\documentclass[aps,prb,twocolumn,showpacs,superscriptaddress,groupedaddress]{revtex4}  
\usepackage{graphicx}  
\usepackage[caption=false]{subfig}
\usepackage{dcolumn}   
\usepackage{bm}        
\usepackage{amssymb}   
\usepackage{amsmath}

\begin{document}



\title{Structural and electronic properties of the incommensurate host-guest Bi-III phase}
\author{D. Kartoon} 
\affiliation{Materials Engineering Department, Ben-Gurion University of the Negev, Beer Sheva 84105, Israel} 
\affiliation{NRCN-Nuclear Research Center Negev, Beer Sheva IL 84190, Israel}

\author{G. Makov} 
\email{makovg@bgu.ac.il}
\affiliation{Materials Engineering Department, Ben-Gurion University of the Negev, Beer Sheva 84105, Israel}

\date{\today}

\begin{abstract}
At high pressure, bismuth acquires a complex incommensurate host-guest structure, only recently discovered. Characterizing the structure and properties of this incommensurate phase from first principles is challenging owing to its non-periodic nature. In this study we use large scale DFT calculations to model commensurate approximants to the Bi-III phase, and in particular to describe the atomic modulations with respect to their ideal positions, shown here to strongly affect the electronic structure of the lattice and its stability. The equation of state and range of stability of Bi-III are reproduced in excellent agreement with experiment using a fully relativistic model. We demonstrate the importance of employing large unit-cells for the accurate description of the geometric and electronic configuration of Bi-III. In contrast, accurate description of the equation of state of bismuth is found to be primarily sensitive to the choice of pseudopotential and exchange-correlation function, while almost completely insensitive to the commensurate approximation.    
\end{abstract}

\maketitle

\section{\label{sec:level1}Introduction}
Incommensurate host-guest structure was first found to exist in elements only twenty years ago, in barium~\citep{Nelmes1999}, surprising condensed matter physicists~\citep{Heine2000}. Since then, more materials were found to acquire such phases at high pressures, including all column-V elements, the alkali metals and some of the alkaline-earths~\citep{Degtyareva2010}. The complex structure of the incommensurate phases is often accompanied by unique electronic properties such as strong coupling superconductivity~\citep{Brown2018} and electride formation~\citep{Woolman2018}, which make them a focus of interest despite the experimental and computational challenges.  

High-pressure research of elemental bismuth has revealed a variety of crystallographic structures, starting with the well-known rhombohedral A7-type structure at ambient pressure and leading to a body centered cubic phase at high pressures~\citep{Degtyareva2004}:
Bi-I (A7, $\textit{h}R2$) $\xrightarrow{2.55\:\textrm{GPa}}$ Bi-II ($\textit{m}C4$) $\xrightarrow{2.7\:\textrm{GPa}}$ Bi-III $\xrightarrow{7.7\:\textrm{GPa}}$ Bi-V (bcc) $<$ 220 GPa.
Whereas the first two low-pressure phases of bismuth have been identified and experimentally characterized in detail very early on~\citep{Bridgman1935,Bundy1958,Klement63,Brugger67,Homan1975}, the nature and even the number of the crystallographic phases comprising Bi-III remained a subject of debate for a long time, due to the complex diffraction patterns obtained at pressures exceeding the second phase transition. 

\begin{figure}
\includegraphics[width=1.0\linewidth]{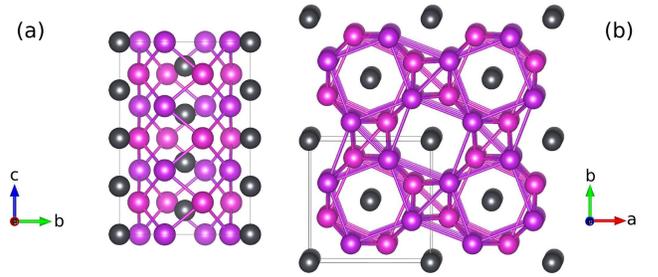}
\caption{\label{fig:Bi-III_phase} Self-hosting Bi-III structure, shown in projection along the $a$-axis (a) and the $c$-axis (b). Host atoms (purple) occupy the positions $8h (x,x+\frac{1}{2},0)$ of space group $I4/mcm $ (140). Guest atoms (black) occupy the positions $2a (0 0 0)$ of space group $I4/mmm$ (139).}
\end{figure}

A model for Bi-III was proposed by Chen \textit{et al.}~\cite{Chen1996}, who studied bismuth at high pressures using synchrotron X-ray diffraction, and argued that Bi-III has a distorted body-centered cubic lattice. This model was in good agreement with the diffraction data, however it resulted in an unphysical volume increase of $ 2\% $ at the Bi-II$\rightarrow$Bi-III transition. Four years later, McMahon \textit{et al.}~\citep{McMahon2000} published their groundbreaking paper, determining that Bi-III, as well as Sb-II, have a complex incommensurate host-guest structure, which is stable across a wide pressure range. In this structure the atoms are arranged in two intertwined sub-lattices (see Fig.~\ref{fig:Bi-III_phase}), sharing the same $a$- and $b$-axes, but with different $c$-axis dimensions. The ratio $c_H/c_G$ between the host's $c$-axis and the guest's is irrational, thus making the crystal non-periodic in its c-direction. Whereas incommensurate structures have been widely known to exist in compound materials for many years~\citep{VanSmaalen}, the concept of self-hosting elements was not introduced until a year earlier for barium~\cite{Nelmes1999}.  

DFT calculations of the incommensurate Bi-III phase were reported by H\"aussermann \textit{et al.}~\citep{Haussermann2002}, who first suggested to use a 32-atom supercell approximant of the Bi-III lattice. This cell contains three host unit-cells and four guest unit-cells, giving a $c_H/c_G$ ratio of 4/3, not very far from the experimentally measured value of $\sim1.31$. This approximant yielded reasonably good agreements to experimental values of density, but a very broad pressure stability range for Bi-III, almost twice its experimental value. Nevertheless, this approximant became the standard numerical scheme for calculating Bi-III, used also in more recent studies calculating phonon spectra and superconductivity transitions~\citep{Brown2018,Khasanov2018,Rodriguez2019}. Although very useful and computationally reasonable, the quality of this commensurate approximation has not been established, and the sensitivity of the various physical properties of the material to it have not been systematically studied.

More recently a synchrotron study by McMahon \textit{et al.}~\citep{McMahon2007} revealed that the positions of the atoms of Bi-III are slightly modulated compared to their host-guest ideal locations. In contrast to the periodicities of the two sub-lattices, which are un-correlated by definition in an incommensurate structure, the modulation of each sub-lattice was found to have strong correlation with the other sub-lattice. This cross influence suggests that these modulations play a crucial role in stabilizing the Bi-III structure. First computational indications supporting the existence of such modulations were obtained using the 4/3 approximation~\citep{Haussermann2002}, but a full description of the modulated structure is unattainable with such small supercell.

Being a heavy element, bismuth is known to have strong relativistic effects which must be included when attempting to calculate its physical properties. Many studies have demonstrated the importance of including spin-orbit correlations (SOC) for calculating the electronic structure of rhombohedral Bi, especially in the vicinity of the Fermi surface~\citep{Gonze1988}, its phonon band structure~\citep{Diaz-Sanchez2007}, its thermodynamic properties~\citep{Diaz-Sanchez2007-PRL} and elastic constants~\citep{Wu2018}. These studies may indicate the importance of SOC for calculating the Bi-III electronic and thermodynamic properties.               

In this study we investigate the Bi-III incommensurate host-guest structure using ab-initio calculations. We employ a series of commensurate approximations to determine the optimal ratio of $c_H/c_G$ and its pressure dependence. The modulations of the atoms with respect to their ideal positions are calculated in large supercells, and compared to experiments. We also investigate the effect of spin-orbit coupling on the electronic structure and the equation of state of Bi-III by using both the scalar-relativistic approximation and fully-relativistic pseudopotential, and compare the results to experiments.

\section{\label{sec:level1}Methods}
The total energy and electronic structure were calculated within the Born-Oppenheimer approximation in the DFT formulation, using a pseudopotential plane-waves method as implemented in the Quantum Espresso package~\cite{QE}. Three different pseudopotentials were employed as listed in table \ref{tab:Pseudos}, two of them scalar-relativistic and the third (PP3) a fully relativistic pseudopotential derived from an atomic Dirac-like equation, all with 15 valence electrons Bi(5d$^{10}$,6s$^{2}$,6p$^{3}$). For the fully relativistic pseudopotential, the electronic structure calculations included spin-orbit coupling. Plane-wave cutoffs were taken to be 40 and 45 Ryd for the scalar-relativistic and the fully-relativistic pseudopotentials, respectively, and gaussian smearing of 0.01 Rydberg width was used. The Brillouin zone was sampled using a regular Monkhorst-Pack grid~\cite{MonkhorstPack} with k-point separation of 0.01 {\AA}$^{-1}$, well within the convergence limit.

\begin{table}[]
\caption{\label{tab:Pseudos}Details of the four pseudopotentials used in this work.}
\begin{tabular}{lllll}
\hline
           & \textbf{PP Type} & \textbf{XC} & \textbf{Rel} & \textbf{Source}                              \\ \hline
\textbf{PP1} & Ultrasoft        & PBE         & Scalar       & GBRV~\cite{GBRV}       \\
\textbf{PP2} & Ultrasoft        & LDA         & Scalar       & GBRV~\cite{GBRV}       \\
\textbf{PP3} & Ultrasoft        & LDA         & Full         & PSLib~\cite{PSlibrary} \\ \hline
\end{tabular}
\end{table}  
  
All four solid phases of bismuth known to exist at ambient temperatures were studied\cite{Degtyareva2004}: The rhombohedral phase with two atoms per unit cell stable at low pressures~\cite{Klement63} (Bi-I), the base-centered monoclinic phase with four atoms in the unit cell stable within a very narrow pressure range~\cite{Brugger67} (Bi-II), the incommensurate host-guest Bi-III structure predicted to be stable between 2.7-7.7 GPa\cite{McMahon2000}, and the high-pressure BCC phase (Bi-V), measured to be stable up to 222 GPa~\citep{Akahama2002}. The Bi-III phase was calculated using a commensurate tetragonal supercell approximant consisting of $n_H$ tetragonal host cells, each containing 8 atoms, and $n_G$ 2-atom BCT guest cells, all stacked along their mutual c-axis, as shown in Fig.~\ref{fig:Bi-III_phase}. As the intertwined host and guest cells share the same supercell, the ratio $n_G/n_H$ is equal to the inverse ratio of their respective unit-cell lengths $c_H/c_G$. To find the best approximation to the incommensurate $c_H/c_G$ ratio, a series of calculations with varying number of guest and host unit cells were made, as described in Table ~\ref{tab:CommAproxTab}. As the experimental $c_H/c_G$ value was found to be close to 1.31~\citep{McMahon2000}, commensurate approximations between 1.2 and 1.4 were calculated, within computational limitations. 

\begin{table}[]
\caption{\label{tab:CommAproxTab}Supercells used to approximate $c_H/c_G$ value.}
\begin{tabular}{cccc}
\hline
\textbf{$n_G$} & \textbf{$n_H$} & \textbf{$c_H/c_G$} & \textbf{\textit{n} atoms}                               \\ \hline
6        & 5         & 1.2       & 52       \\
5        & 4         & 1.25      & 42       \\
13       & 10        & 1.3       & 106      \\
33       & 25        & 1.32      & 266      \\
4        & 3         & 1.3333    & 32       \\
7        & 5         & 1.4       & 54       \\ \hline
\end{tabular}
\end{table}  

Following the setting of the ideal atomic positions described above, optimization of the supercell atomic configurations were performed separately at each target pressure between 3-8 GPa. The Broyden-Fletcher-Goldfarb-Shanno (BFGS) algorithm~\citep{BFGS} was used to achieve the optimized volumes and atomic locations for each pressure. The relaxation process resulted in optimized internal energies $E(V)$ for each set of calculations, to which a Birch-Murnaghan equation-of-state~\cite{BirchMurnaghan} was fitted. The pressure P(V) and enthalpy H(V) were obtained as derivatives of the total energy. Independent sets of calculations were performed for each of the four phases of bismuth.   

\section{\label{sec:level1}Results and discussion}
\subsection{Unit-cell optimization}
\begin{figure}
\includegraphics[width=1.0\linewidth]{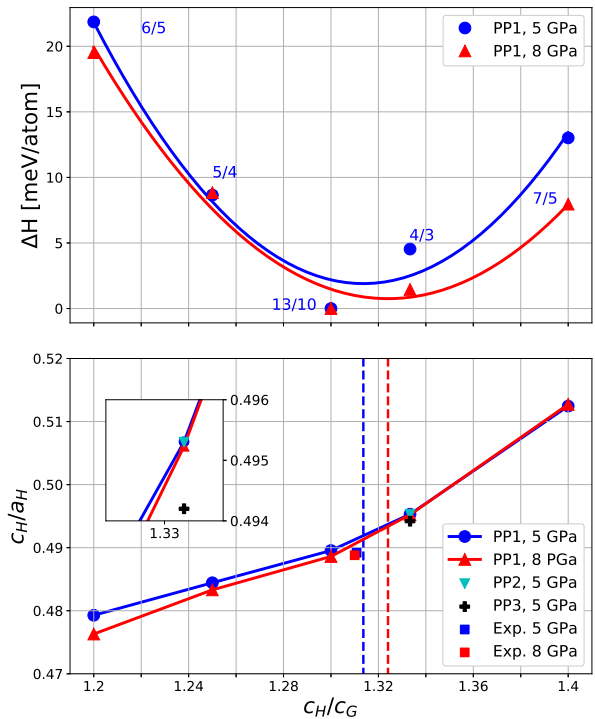}
\caption{\label{fig:optim_ratio} Upper panel: relative enthalpies calculated for Bi-III at 5 GPa (blue) and 8 GPa (red) using several commensurate approximations. DFT calculations results are marked with symbols and labeled with their respective $n_G/n_H$. A parabolic fit (solid line) gives the optimal value of $c_H/c_G$. Lower panel: optimized aspect ratios of the host cell $c_H/a_H$ for the different commensurate approximations. Optimal values of $c_H/c_G$ at 5 and 8 GPa are marked by vertical dashed lines. Inset in lower panel shows a closeup around the commensurate approximation $c_H/c_G=4/3$. All calculations using PBE scalar-relativistic pseudopotential (PP1), except for the colored symbols on the lower panel using PP2 (cyan) and PP3 (black). Experimental values are taken from Degtyareva \textit{el al.}~\cite{Degtyareva2004} }
\end{figure}

Five supercells with different commensurate ratios between 1.2 and 1.4, as detailed in Table ~\ref{tab:CommAproxTab}, were considered to find the best approximation to the irrational $c_H/c_G$ value at two pressures $P$=5 GPa and $P$=8 GPa. The relative enthalpies obtained from the calculations at the two pressures using pseudopotential PP1 (Table \ref{tab:Pseudos}) are plotted in Fig.~\ref{fig:optim_ratio}. As the 33/25 supercell calculation proved too demanding to be performed with the same k-space resolution as the other calculations, its relative enthalpy is not displayed in the figure. A parabolic fit to the results presents a minimum enthalpy at $c_H/c_G=1.313$ at $P$=5 GPa and $c_H/c_G=1.324$ at $P$=8 GPa, very close to the experimental values of 1.311 and 1.310, respectively~\citep{Degtyareva2004}. Although the calculated volume per-atom at a given pressure does not depend on the commensurate approximation (less than $0.25\%$), the aspect-ratio of the host unit-cell $c_H/a_H$ is affected by it, as demonstrated in the lower panel of Fig.~\ref{fig:optim_ratio}. The value of $c_H/a_H=0.491$ at the optimal commensurate ratio agrees well with the experimental value, and similarly does not depend significantly on pressure~\citep{Degtyareva2004}.

Due to their computational cost, calculations with a fully-relativistic pseudopotential (PP3 in Table \ref{tab:Pseudos}) were performed only with the 4/3 supercell, which contains 32 atoms. The $c_H/a_H$ ratio obtained with various pseudopotentials (shown in inset of the lower panel of Fig.~\ref{fig:optim_ratio}) demonstrates very weak sensitivity to relativistic effects or the choice of exchange-correlation functional, despite the fact the the equilibrium volumes of these calculations show significant variance, as will be shown later.  

The stabilizing mechanism of the incommensurate structure is not entirely clear. Strong evidence for such a mechanism was suggested by McMahon \textit{el al.}~\citep{McMahon2007}, who used synchrotron radiation to obtain high resolution diffraction images of single-crystal Bi-III. Their study revealed that the bismuth atoms are slightly displaced from their ideal host-guest positions (depicted in Fig.~\ref{fig:Bi-III_phase}), and that the displacements of each sub-lattice have a periodicity related to that of the other sub-lattice. Such intermodulation,  possible only in the framework of a 4-dimensional superspace~\citep{VanSmaalen}, were also reported for other crystalline materials~\citep{Toudic2008}, and suggested to play a crucial role in the stability and phase transitions of such structures. 

\begin{figure}
\includegraphics[width=1.0\linewidth,scale=1]{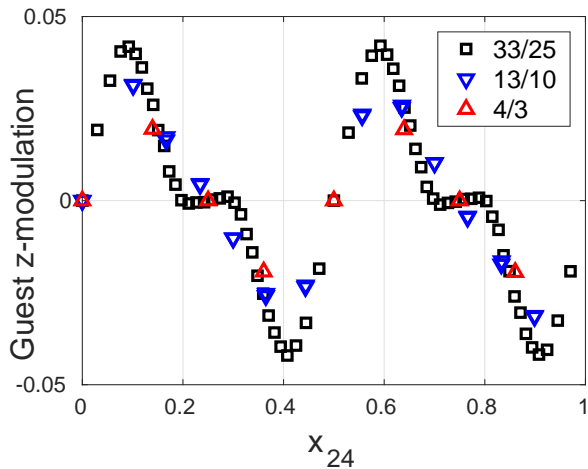}
\caption{\label{fig:GuestZmod} Modulations of the guest atoms along the z-axis with respect to their ideal positions as a function of $x_{24}$. The modulations are given in the guest lattice units. $x_{24}$ represents the z-position of the guest atom in host lattice units, and defined as $x_{24}=Z(\text{Guest})\cdot c_G/c_H$. Modulations were taken from 33/25 (black), 13/10 (blue) and 4/3 (red) supercell calcuations at $P$=5 GPa.}
\end{figure}

Prior first-principles calculations of Bi-III were carried out using 32-atom supercells  containing three host cells and four guest cells~\citep{Haussermann2002,Haussermann2004,Chen_2016,Brown2018,Khasanov2018}. This approximation  proves to be very useful for calculating many of the material's properties, however the limited number of atoms in each sub-lattice makes it inappropriate for investigating higher-order variations such as the modulations of the atomic positions. Using a variable-cell relaxation process with supercells containing up to 266 atoms, we were able to explore these modulations theoretically for the first time, and compare them to the experimental results.
  
An indication of the significance of the modulations in stabilizing the incommensurate cell arises from their contribution to reducing the total enthalpy of the system. In the 13/10 supercell, for example, the modulations lower the average total energy of each atom by $\sim$20 meV, comparable to the energy differences along the whole commensurate span 1.2-1.4, as presented in the upper panel of Fig.~\ref{fig:optim_ratio}. The most pronounced deviation of the atoms from the ideal host-guest structure, is the pairing of the guest atoms which are aligned in one-dimensional chains surrounded by rectangular rings of the host atoms (see Fig.~\ref{fig:Bi-III_phase}). This pairing reduces the distance between neighboring guest atoms from approximately 3.2 {\AA} to 3.07 {\AA}, very close to its experimental values obtained in Bi-III at the same pressure~\citep{McMahon2007} and in the A7 phase at ambient pressure~\citep{Cucka1962}. 

Although the pairing itself is evident even in the 32-atom supercell~\citep{Haussermann2002}, full description of the complex periodic behavior of the modulation along the z-axis,as observed experimentally~\citep{McMahon2007}, requires much larger supercells, as demonstrated in Fig.~\ref{fig:GuestZmod}. The calculated z-modulations of the guest atoms are plotted as a function of their relative position in the host unit cell, all "folded" into one representative cell (same coordinates were used by McMahon \textit{et al.}~\citep{McMahon2007}, detailed definition can be found therein). In these coordinates, the host atoms are located at positions $x_{24}=0,0.5,1$. The saw-tooth distribution of the guest modulations in these coordinates, apparent in both experiments and calculations, indicates that the maximum displacement occurs very close to the planes of the host's rectangular rings ($z=0,\tfrac{1}{2}$ in the host unit cell), and these displacements decrease non-monotonically between the planes. Increasing the supercell size to 33/25 (266 atoms) exposes kinks in the modulation occurring at approximately $x_{24}=0.25$ and $x_{24}=0.75$, which probably split into double-kinks in the experiments. 

\begin{figure}
\includegraphics[width=1.0\linewidth,scale=1]{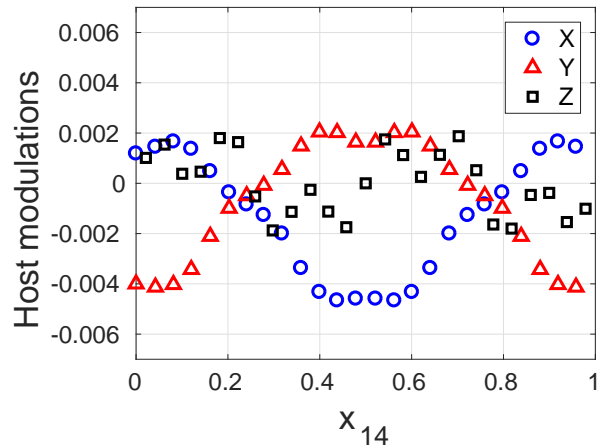}
\caption{\label{fig:HostXYZmod} Modulations of the host atoms along the x (blue),y (red) and z (black) axes in respect to their ideal positions as a function of $x_{14}$. The modulations are given in the host lattice units. $x_{14}$ represents the z-position of the host atom in the guest lattice units, and defined as $x_{14}=Z(\text{Host})\cdot c_H/c_G$. Modulations were taken from the 33/25 supercell calculation at $P$=5 GPa.}
\end{figure}

\begin{figure}
\includegraphics[width=1.0\linewidth,scale=1]{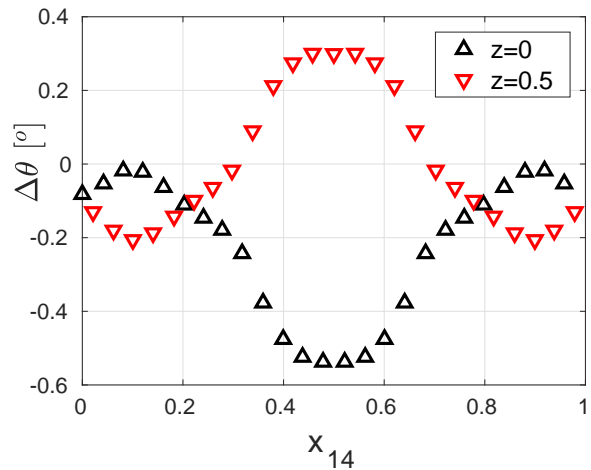}
\caption{\label{fig:HostThetamod} $\theta$-modulations of the host atoms in respect to their ideal positions as a function of $x_{14}$. $\theta$ is the rotation of the host's rectangular rings in the $xy$-plane given in degrees. The rotation of the z=0.5 atoms ($ \bigtriangledown $) is shifted by its mean value ($\theta \cong 3^{\circ}$) for clarity. $x_{14}$ represents the z-position of the host atom in the guest lattice units, and defined as $x_{14}=Z(\text{Host})\cdot c_H/c_G$. Modulations were taken from the 33/25 supercell calculation at $P$=5 GPa.}
\end{figure}

In contrast to the guest atoms, which, by restriction of the superspace group symmetry, can only shift in the $z$-direction, the host atoms can be displaced in all three dimensions of the tetragonal lattice. The modulations of the host atoms are much smaller than those of the guest atoms, and occur mainly in the $xy$-plane, as can be seen in Fig~\ref{fig:HostXYZmod}. The rectangular rings of the host rotate alternately to form two-dimensional zigzag chains along the $z$-axis; thus reducing the shortest distance between adjacent host atoms from 3.23-3.3 {\AA} to 3.17 {\AA} on average, as also obtained by H\"aussermann \textit{et al.}~\citep{Haussermann2002}. However, our more detailed calculations reveal that these rotations are influenced by the guests periodicity,in accordance with experiments, as demonstrated in Fig.~\ref{fig:HostThetamod}.

\subsection{Electronic configuration}

\begin{figure}
\includegraphics[width=1.0\linewidth,scale=1]{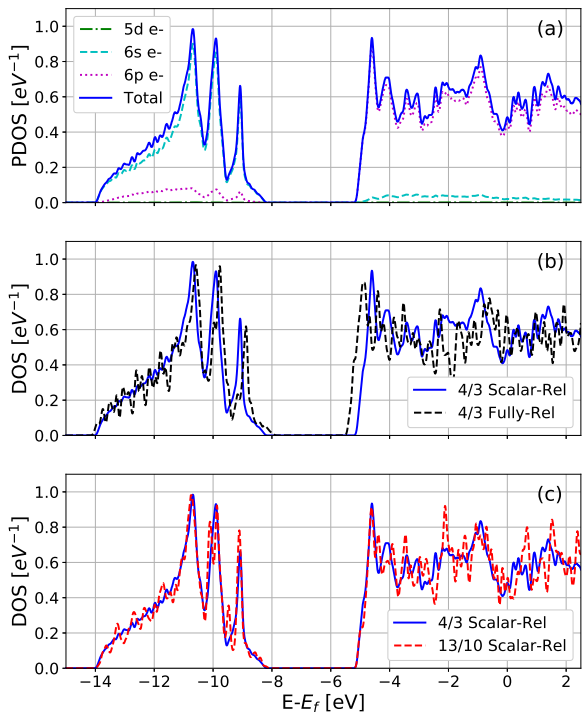}
\caption{\label{fig:Bi-III_PDOS} Bi-III density of states (DOS) and its orbital projections (PDOS) at 5 GPa (a) using scalar-relativistic PBE pseudopotential (PP1). Total DOS is compared to fully-relativistic (PP3) pseudopotential with the same commensurate approximant 4/3 (b), and to 13/10 approximant (c) using the same pseudopotential (PP1). }
\end{figure}

The electronic structure of the Bi-III phase was investigated to gain additional insight into the stabilizing mechanism of the incommensurate host-guest structure. The electronic density of states (DOS) of Bi-III in the 4/3 approximation was calculated with both scalar-relativistic (PP1) and fully-relativistic (PP3) pseudopotentials, along with its orbital projections, as illustrated in Fig.~\ref{fig:Bi-III_PDOS}. It is apparent that the s-band and the p-band are distinct even at elevated pressure ($P$=5 GPa). The s-band has a unique triple-peaked shape, typical to Bi-III structure calculated with the 4/3 approximant~\citep{Haussermann2002,Brown2018}. Studying the topology of the integrated local DOS (ILDOS) at the energy range of each of the s-band peaks, reveals that the first peak corresponds to the bonds between the guest atoms along the one-dimensional chain (distance $\simeq$3.07 {\AA}), whereas the other two peaks correspond to the bonds within the host's zigzag chains, with two close distances - 3.38 {\AA} (second peak) and 3.41 {\AA} (third peak). It is interesting to note that in contrast to the first peak, the second and third peaks depend on the details of the atomic structure and hence split when using other approximants, as demonstrated in Fig.~\ref{fig:Bi-III_PDOS}c. The DOS exhibits a distinct, although mild, drop in the vicinity of the Fermi level, reminiscent of Bi semi-metallic nature at ambient pressure. Relativistic effects have little influence on the DOS, mostly making it more metallic, which agrees well with previous work done with the same commensurate approximation~\citep{Haussermann2002,Brown2018}.

\begin{figure}
\includegraphics[width=1.0\linewidth,scale=1]{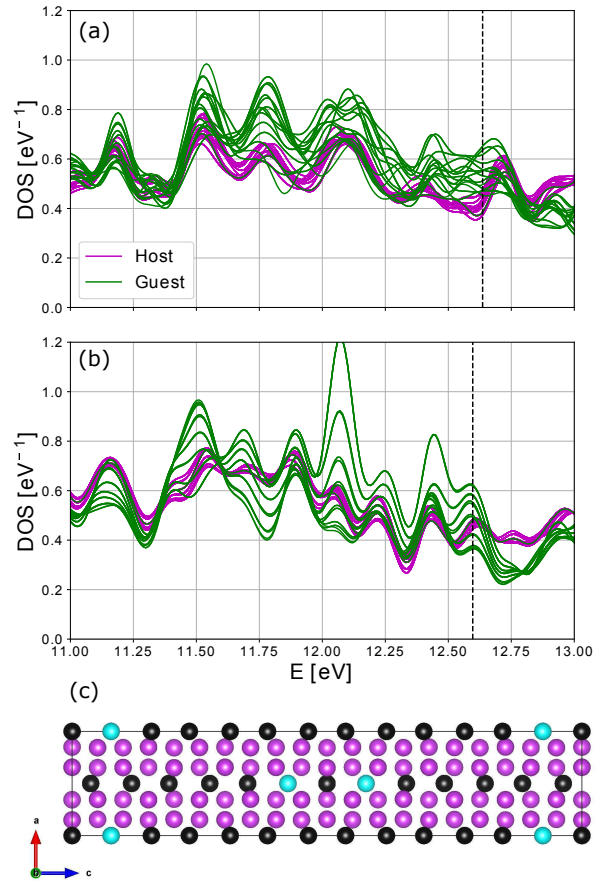}
\caption{\label{fig:HG_PDOS} Projections of the Bi-III density of states (DOS) on the host (magenta) and guest (green) p-orbital states. Calculations using 13/10 commensurate approximation with ideal (a) and modulated (b) atomic positions. The Fermi energy is marked by a vertical dashed line. The host (magenta) and guest (black) atoms as projected along the b-axis of the unit cell are illustrated in (c).}
\end{figure}

Due to the intricate structure of Bi-III, it is interesting to compare the electronic structure of the atoms comprising the two intertwined sub-lattices. Previous studies~\citep{Haussermann2002,Khasanov2018} found no difference between the density of states projected on the sites of the guest and of the host atoms. However, making this comparison with the larger 13/10 supercell reveals distinct differences between the host and the guest atoms, as shown in Fig.~\ref{fig:HG_PDOS}. Whereas the host atoms p-projected DOS are relatively similar to each other, the guest atoms demonstrate wide distribution throughout the p-band, varying both from the host atoms and from each other. Moreover, it appears that these variations are coupled to the modulations of the atoms from their ideal positions, and that these modulations increase them dramatically. The relations between the DOS dispersion of the guest atoms and the modulations can also be explored by identifying each projected DOS with its spatial location. For example, all of the projected DOS with the largest variations compared to the mean DOS (highest peaks at $E$=12.1 eV in Fig.~\ref{fig:HG_PDOS}b) are located \textit{farthest} from the host atoms (highlighted in cyan in Fig.~\ref{fig:HG_PDOS}c), which means that these atoms have the \textit{smallest} spatial modulations ($ x_{24}=0.25,0.75 $ in the coordinates of Fig.~\ref{fig:GuestZmod}). This implies that the modulations have non-local effects on the electronic structure of the lattice. 

\subsection{Equation of state}
\begin{figure}
\includegraphics[width=1.0\linewidth]{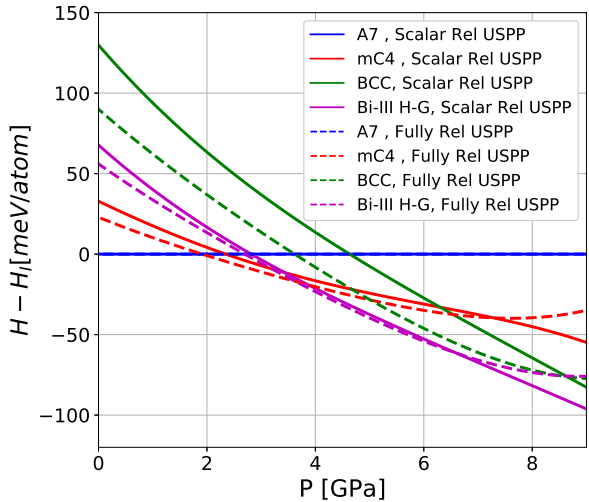}
\caption{\label{fig:Enthalpy} Relative enthalpies of the monoclinic Bi-II (red), incommensurate host-guest Bi-III (magenta) and BCC (green) phases in respect to the rhombohedral A7 ambient phase(blue). Solid (dashed) lines represent calculations using scalar (fully) relativistic pseudopotential with PBE (LDA) exchange-correlation (PP1 and PP3 in Table ~\ref{tab:Pseudos}). }
\end{figure}

At low-temperatures bismuth is known to adopt one of four crystallographic structures: rhombohedral with two atoms per unit cell, base-centered monoclinic with four atoms in the unit cell, incommensurate host-guest or BCC, depending on pressure~\cite{Degtyareva2004}. The total energies for all four phases were calculated at several molar volumes and fitted with a Birch-Murnaghan equation-of-state~\cite{BirchMurnaghan} $E(V)$. The pressure $P(V)$ and enthalpy $H(V)$  were obtained by derivation. Enthalpy differences with respect to the A7 phase are presented in Fig.~\ref{fig:Enthalpy} as a function of pressure for Bi-II, Bi-III and BCC phases, calculated with scalar-relativistic and fully-relativistic pseudopotentials (PP1 and PP3 from Table~\ref{tab:Pseudos}, respectively). Calculations using either pseudopotential predict phase Bi-II to be stable in a pressure range of about 1 GPa, starting at $P$=2.3 GPa (PP1) or $P$=1.9 GPa (PP3). This result is slightly shifted compared to the experimental phase diagram at ambient conditions~\citep{Degtyareva2004}, but is inconsistent with previous low-temperature experiments~\cite{Compy1970,Homan1975} which reported Bi-II to vanish below 160K, and previous calculations~\citep{Haussermann2002}, which found Bi-II to be thermodynamically unstable at zero temperature, resulting in a direct transition from A7 to the Bi-III phase. However, there is experimental evidence of a Bi-II$\rightarrow$Bi-III phase transition at pressures of about 3.6 GPa and temperatures as low as 77K~\citep{Mori1971},  with good agreement with our calculations. The discrepancies between the experimental results at low temperatures have not yet been resolved. 

\begin{figure}
\includegraphics[width=1.0\linewidth,scale=1]{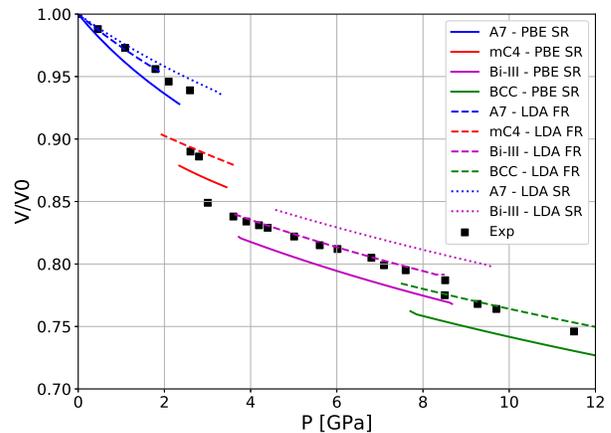}
\caption{\label{fig:Bi_Compress} Compressibility of bismuth at pressures up to 12 GPa. Calculated phases include A7 (blue), monoclinic (red), Bi-III (magenta) and BCC (green), with PBE scalar-relativistic pseudopotential (solid lines) and LDA fully-relativistic pseudopotentials (dashed). Experimental data is taken from Degtyareva \textit{et. al}~\citep{Degtyareva2004}}
\end{figure}

The transition from Bi-III to BCC proves to be highly sensitive to the choice of pseudopotential. The fully-relativistic pseudopotential with LDA exchange-correlation produces a transition pressure of $P$=8.5 GPa, very close to the experimental values of $P$=7.7 GPa~\citep{Aoki1982} and $P$=8 GPa~\citep{McMahon2000}. In contrast, the scalar-relativistic pseudopotential with PBE exchange-correlation yields a much higher transition pressure $P$=13 GPa (outside the scale of Fig.~\ref{fig:Enthalpy}), very similar to the transition pressure obtained in previous calculations~\citep{Haussermann2002}. This discrepancy of the stability range of the incommensurate host-guest phase indicates the importance of the spin-orbit coupling to the proper description of bismuth equation of state. 

\begin{figure}
\includegraphics[width=1.0\linewidth,scale=1]{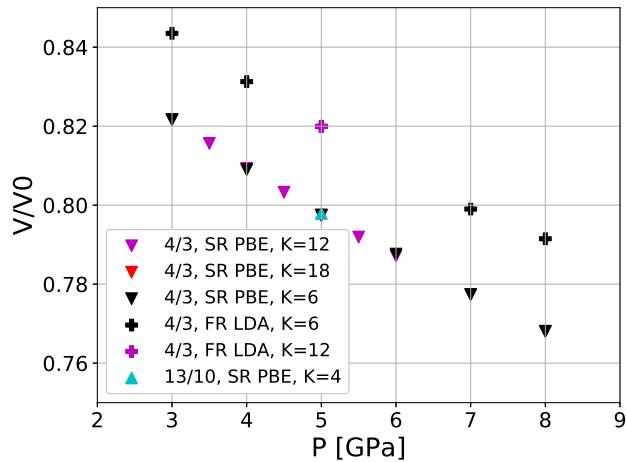}
\caption{\label{fig:Bi_Compress_sesitivity} Compressibility of Bi-III at various pressures, calculated with pseudopotentials PP1 ($ \blacktriangledown $) and PP3 (+) from Table~\ref{tab:Pseudos} using the 4/3 and the 13/10 ($ \blacktriangle $) commensurate approximations. Different k-point sampling densities are drawn with different colors.} 
\end{figure}

To further examine the relativistic contribution, we used both fully-relativistic (PP3) and scalar-relativistic pseudopotentials (PP1 and PP2) to calculate the pressure equation of state of bismuth, as shown in Fig.~\ref{fig:Bi_Compress}. The equation of state was found to be well-converged and insensitive to the commensurate approximation, as demonstrated in Fig.~\ref{fig:Bi_Compress_sesitivity}. The fully-relativistic pseudopotential shows excellent agreement with the experimental results for all four phases examined, compared to the other potentials. Some of the difference in the results between the two main potentials used in this work (PP1 and PP3) is expected to arise from their different exchange-correlation functional; the fully-relativistic pseudopotential employs the LDA exchange-correlation, known to be overbinding for nearly all materials~\cite{Haas2009}. To test this effect we calculated the pressure volume curve of Bi-III also with a scalar relativistic LDA pseudopotential (PP2), and found it to exhibit slightly different $P(V)$ behavior, as shown in Fig.~\ref{fig:Bi_Compress}. Spin-orbit correlations have been found to be of crucial important for accurate description of the electronic configuration~\citep{Gonze1988} and phonon spectrum~\citep{Diaz-Sanchez2007} of A7 bismuth. These results suggest that the spin-orbit corrections for bismuth are also important for calculating the Bi-III equation-of-state.  

As previously mentioned, the calculated density does not depend on the precise incommensurate value of $c_H/c_G$. Fig.~\ref{fig:Bi_Compress_sesitivity} demonstrates this insensitivity, along with the dependence on the pseudopotential and the exchange-correlation function.   
 
\section{\label{sec:level1}Conclusions}
The incommensurate host-guest phase Bi-III was investigated using DFT approach. A series of commensurate approximants were used to describe the aperiodic lattice, producing the theoretical predictions for $c_H/c_G$ at two pressures in good agreement with experimentally determined values. Modulations of the atoms in respect to their ideal locations were calculated in detail and were shown to contribute significantly to structure stabilization. It was revealed that although each of the sub-lattices has a different periodicity, the periodicity of the modulations in each of the sub-lattices is correlated to the periodicity of the other. Calculated modulations of both the guest and the host atoms reproduce the experimental results with very good accuracy. 

The electronic structure of Bi-III was investigated, and found to be only mildly affected by spin-orbit correlation effects. Non-negligible differences in the projected density of states were found between the host and the guest atoms, which demonstrate substantial scatter. This scatter is correlated to the host atoms periodicity, similar to the modulations of the guest atoms, and found to be enhanced by them. These findings emphasize the key role of the inner modulations within the incommensurate host-guest structures in enabling long-distance correlations between the atoms of each sub-lattice.   

The stability range of Bi-III, as well as the equation of state of bismuth, was calculated using different pseudopotentials and exchange-correlation functions. Excellent agreement with experiment was achieved, but only when employing the fully relativistic pseudopotential. In contrast to the geometric optimization parameters, which demonstrate relatively strong dependence on the commensurate approximant but are not affected by the choice of pseudopotential, the equation-of-state and its derivatives show the reverse behavior, as they are sensitive both to the choice of pseudopotential and to the exchange-correlation function, but not to the exact $c_H/c_G$ value.

These findings offer a new insight into the interactions between the atoms comprising of the two sub-lattices of the Bi-III structure, helping to understand the origin of its stability and to define the regimes of its computational sensitivity.

\begin{acknowledgments}
Computational support was provided by the NegevHPC project~\cite{negevhpc}.
\end{acknowledgments}
\bibliographystyle{apsrev4-1} 
\bibliography{BiIII_refs}

\end{document}